\documentstyle[psfig]{mn}

\newif\ifAMStwofonts
\ifoldfss
  \ifCUPmtlplainloaded \else
    \NewTextAlphabet{textbfit} {cmbxti10} {}
    \NewTextAlphabet{textbfss} {cmssbx10} {}
    \NewMathAlphabet{mathbfit} {cmbxti10} {} % for math mode
    \NewMathAlphabet{mathbfss} {cmssbx10} {} %  "   "    "
  \fi
  \ifAMStwofonts
    \ifCUPmtlplainloaded \else
      \NewSymbolFont{upmath} {eurm10}
      \NewSymbolFont{AMSa} {msam10}
      \NewMathSymbol{\upi}     {0}{upmath}{19}
      \NewMathSymbol{\umu}     {0}{upmath}{16}
      \NewMathSymbol{\upartial}{0}{upmath}{40}
      \NewMathSymbol{\leqslant}{3}{AMSa}{36}
      \NewMathSymbol{\geqslant}{3}{AMSa}{3E}

    \fi
  \fi
\fi % End of OFSS

\ifnfssone
  \newmathalphabet{\mathit}
  \addtoversion{normal}{\mathit}{cmr}{m}{it}
  \addtoversion{bold}{\mathit}{cmr}{bx}{it}
  \newmathalphabet{\mathbfit} % math mode version of \textbfit{..}
  \addtoversion{normal}{\mathbfit}{cmr}{bx}{it}
  \addtoversion{bold}{\mathbfit}{cmr}{bx}{it}
  \newmathalphabet{\mathbfss} % math mode version of \textbfss{..}
  \addtoversion{normal}{\mathbfss}{cmss}{bx}{n}
  \addtoversion{bold}{\mathbfss}{cmss}{bx}{n}
  \ifAMStwofonts
    \ifCUPmtlplainloaded \else
      \UseAMStwoboldmath
      \makeatletter
      \new@mathgroup\upmath@group
      \define@mathgroup\mv@normal\upmath@group{eur}{m}{n}
      \define@mathgroup\mv@bold\upmath@group{eur}{b}{n}
      \edef\UPM{\hexnumber\upmath@group}
      \new@mathgroup\amsa@group
      \define@mathgroup\mv@normal\amsa@group{msa}{m}{n}
      \define@mathgroup\mv@bold\amsa@group{msa}{m}{n}
      \edef\AMSa{\hexnumber\amsa@group}
      \makeatother
      \mathchardef\upi="0\UPM19
      \mathchardef\umu="0\UPM16
      \mathchardef\upartial="0\UPM40
      \mathchardef\leqslant="3\AMSa36
      \mathchardef\geqslant="3\AMSa3E
    \fi
  \fi
\fi % End of NFSS release 1

\ifnfsstwo
  \DeclareMathAlphabet{\mathbfit}{OT1}{cmr}{bx}{it}
  \SetMathAlphabet\mathbfit{bold}{OT1}{cmr}{bx}{it}
  \DeclareMathAlphabet{\mathbfss}{OT1}{cmss}{bx}{n}
  \SetMathAlphabet\mathbfss{bold}{OT1}{cmss}{bx}{n}
  \ifAMStwofonts
    \ifCUPmtlplainloaded \else
      \DeclareSymbolFont{UPM}{U}{eur}{m}{n}
      \SetSymbolFont{UPM}{bold}{U}{eur}{b}{n}
      \DeclareSymbolFont{AMSa}{U}{msa}{m}{n}
      \DeclareMathSymbol{\upi}{0}{UPM}{"19}
      \DeclareMathSymbol{\umu}{0}{UPM}{"16}
      \DeclareMathSymbol{\upartial}{0}{UPM}{"40}
      \DeclareMathSymbol{\leqslant}{3}{AMSa}{"36}
      \DeclareMathSymbol{\geqslant}{3}{AMSa}{"3E}
    \fi
  \fi
\fi % End of NFSS release 2

\ifCUPmtlplainloaded \else
  \ifAMStwofonts \else % If no AMS fonts
    \def\upi{\pi}
    \def\umu{\mu}
    \def\upartial{\partial}
  \fi
\fi

\title[Gravitational lensing in Abell 2219]{Infrared observations of
gravitational lensing in Abell 2219 with CIRSI}
\author[M. E. Gray \etal]
       {Meghan E. Gray$^1$\thanks{email: {\em meg@ast.cam.ac.uk}},
       Richard S. Ellis$^1$, Alexandre Refregier$^1$, Jocelyn
       B\'ezecourt$^2$, \cr Richard G. McMahon$^1,$ Martin
       G. Beckett$^{1,3}$, Craig D. Mackay$^1$,\cr and Michael D.
       Hoenig$^1$\\ 1. Institute of Astronomy, Madingley Road,
       Cambridge CB3 0HA \\ 2. Kapteyn Institute, Postbus 800, 9700 AV
       Groningen, The Netherlands\\ 3. Observatories of the Carnegie
       Institution of Washington, 813 Santa Barbara Street, Pasadena,
       CA 91101, USA}
\date{Accepted
      Received
      in original form }

\pagerange{\pageref{firstpage}--\pageref{lastpage}}
\pubyear{2000}

\def\etal{et al. }
\begin{document}

\maketitle

\label{firstpage}

\begin{abstract}
We present the first detection of a gravitational depletion signal at
near-infrared wavelengths, based on deep panoramic images of the
cluster Abell 2219 ($z$=0.22) taken with the Cambridge Infrared
Survey Instrument (CIRSI) at the prime focus of the 4.2m William
Herschel Telescope. Infrared studies of gravitational depletion offer
a number of advantages over similar techniques applied at optical
wavelengths, and can provide reliable total masses for intermediate
redshift clusters. Using the maximum likelihood technique developed by
Schneider, King \& Erben (1999), we detect the gravitational depletion
at the $3\sigma$ confidence level.  By modeling the mass distribution
as a singular isothermal sphere and ignoring the uncertainty in the
unlensed number counts, we find an Einstein radius of $\theta_{\rm E}
\simeq 13.7^{+3.9}_{-4.2}$ arcsec (66\% confidence limit). This
corresponds to a projected velocity dispersion of $\sigma_v \sim 800$
km s$^{-1}$, in agreement with constraints from strongly-lensed
features. For a Navarro, Frenk and White mass model, the radial
dependence observed indicates a best-fitting halo scale length of
\mbox{125 h$^{-1}$ kpc}. We investigate the uncertainties arising from
the observed fluctuations in the unlensed number counts, and show that
clustering is the dominant source of error. We extend the maximum
likelihood method to include the effect of incompleteness, and discuss
the prospects of further systematic studies of lensing in the
near-infrared band.
\end{abstract}

\begin{keywords}
gravitational lensing -- galaxies: clusters: individual: Abell 2219 --
infrared: galaxies
\end{keywords}

\section{Introduction}

Gravitational lensing is now recognised as a valuable probe of the
mass distribution in intermediate redshift ($z < 0.4$) galaxy clusters
independent of any assumptions about the nature of the lensing
material (see reviews by Fort \& Mellier 1995, Narayan \& Bartelmann
1997, Mellier 1999, Bartelmann \& Schneider 1999). For lenses at such
redshifts, uncertainties arising from the redshift distribution of
background sources are minimized and the angular scales of both weak
and strongly-lensed features are well-suited for precise studies.

On small angular scales in super-critical systems, multiply-lensed
arcs can provide useful absolute mass estimates provided spectroscopic
redshifts and geometric constraints on the location of the various
critical lines are available (e.g. Kneib \etal 1994, 1996). The most
promising progress in constraining the mass on large scales has come
from weak shear measurements (e.g. Fischer 1999; Hoekstra \etal 1998;
Clowe \etal 1998) for which sophisticated inversion techniques have
been developed (e.g. Kaiser \& Squires 1993; Kaiser 1995).

However, cluster mass determinations based on weak shear signals are
not without limitations.  Firstly, the mass reconstruction from weak
shear is non-trivial because of boundary effects due to the finite
field of the data.  To counter this effect, several finite-field
methods have been proposed (e.g. Seitz \& Schneider 1996).  Secondly,
since the shear arises from the {\em gradient} of the gravitational
potential, the mass reconstruction is only known to within an additive
constant.  Accordingly, masses based on shear measurements are subject
to a possible upward correction arising from a `mass sheet
degeneracy'.

With sufficiently wide-field data it is possible to make the
assumption that the surface mass density will approach zero at large
distances from the cluster.  However, with independent knowledge of the
magnification of the lens, the mass-sheet degeneracy can be broken
regardless of the field of view of the data.  Two methods have been
proposed to make use of this magnification information and calibrate
the absolute scale of the mass distribution, either through the change
of image size at fixed surface brightness (Bartelmann \& Narayan 1995)
or source counts (Broadhurst, Taylor \& Peacock 1995).

This paper is concerned with exploring the role that infrared imaging
offers in the gravitational depletion (or `convergence') method for
estimating the total masses of clusters. The depletion method was
first suggested by Broadhurst, Taylor \& Peacock (1995) who predicted
the diminution in background galaxy surface number density as a function of
radius expected behind a lensing cluster. Here we are concerned with
extending the original test to near-infrared wavelengths where, in
principle, there are significant advantages, namely the flatter
number-count slope and a more accurate colour discrimination between
foreground and background populations.  The unique wide-field
capabilities of the panoramic near-infrared Cambridge Infrared Survey
Instrument (CIRSI, Beckett \etal 1998) allow us to test the method
on the rich cluster Abell 2219 ($z$=0.22).

A plan of the paper follows. In $\S$2 we review the gravitational
depletion method illustrating the difficulties associated with its
implementation at optical wavelengths and the potential gains of
repeating the experiment at near-infrared wavelengths. In $\S$3 we
present new observations of Abell 2219 made at the prime focus of the
4.2m William Herschel telescope and discuss the techniques used to
reduce the data as well as the methods used to create a sample of
background galaxies. $\S$4 discusses the depletion signal observed in
the context of various mass models and reviews the uncertainties
involved. In $\S$5 we discuss the prospects of routinely estimating
cluster masses using this method both with CIRSI and with the upcoming
suite of wide field infrared survey telescopes.

\section{Gravitational Depletion}

The gravitational depletion or the `convergence' method of breaking
the mass-sheet degeneracy (Broadhurst, Taylor, \& Peacock 1995) relies
on the change in the surface number density of background galaxies
induced by the magnification effect of a gravitational lens. Since
only source counts are involved, exquisite image quality (essential
for shear measurements) is not necessary.  Furthermore, as the effect
depends on the magnification $\mu$, absolute mass estimates are
possible if the redshift distribution of the background sources is
reasonably well-understood.

The intrinsic (unlensed) counts $n_{0}$ of galaxies brighter than some
limiting magnitude $m$ are transformed to the observed (lensed) counts
$n$ by
\begin{equation} \label{eqn-N}
n(<m) = n_0(<m){\mu}^{2.5\alpha-1},
\end{equation}
where $\alpha$ is the logarithmic slope of the number counts,
\begin{equation}
\label{eq:alpha}
\alpha \equiv \frac{d\log n_{0}(m)}{dm}.
\end{equation}
Two competing effects serve to change the lensed surface number
density of the background galaxies. Source magnification clearly
increases the surface number density by magnifying galaxies that would
otherwise be fainter than the limiting magnitude. However, focusing
within the beam dilutes the overall surface number density. The net
effect depends on the value of $\alpha$: for $\alpha>0.4$ there will
be an overall increase in observed surface number density, while for
$\alpha<0.4$ a depletion is measured.

Unfortunately, at the limits where a sizeable fraction of field
galaxies are expected to be behind an intermediate redshift cluster,
the slope of the optical field counts, \mbox{$\alpha \approx 0.4$}, produces
only a weak effect. In order to demonstrate the effect in one of the
most massive clusters known, Abell 1689 ($z$=0.18), Taylor \etal
(1998) restrict their analysis to a red subsample known {\em a
priori} from blank field studies to have a flatter slope
($\alpha<0.4$). By colour-selecting sources redder than the sequence
of cluster spheroidals, Taylor \etal simultaneously secure a
background population whose $\alpha$ is sufficiently low for the
depletion method to work, and with an unlensed surface number density of 12
arcmin$^{-2}$.  Fort, Mellier \& Dantel-Fort (1997) also search for an
optical depletion effect behind the cluster CL0024+1654, by restricting
their search to the magnitude ranges $26<B<28$ and $24<I<26.5$ where
the slopes are found to be \mbox{$\alpha_{\rm B}=0.17$} and
\mbox{$\alpha_{\rm I} = 0.25$}.  However, they apply no colour
selection to remove faint cluster members in this magnitude range, and
furthermore are left with an extremely low number density of only a
few galaxies per square arcmin.  Similarly, Athreya \etal (in
preparation) use photometric redshifts to isolate a background
population for their weak lensing analysis of MS 1008-1224, but are
hampered by a small field of view and background clustering.  Clearly,
balancing accurate discrimination of the lensing foreground from the
background population while maintaining both a flat slope and
sufficient numbers of background galaxies is of crucial importance,
and it is for this reason that we turn to wide-field infrared observations.
 
Here we are concerned with extending the depletion method to
near-infrared wavelengths, and we illustrate the possible advantages via
an initial application to the rich cluster Abell 2219. The slope of
the number counts flattens at longer wavelengths because of a reduced
sensitivity to intermediate-redshift star forming galaxies (Ellis
1997). Moreover, this flattening occurs at progressively brighter
apparent magnitudes. {\it Without any colour selection}, the slope of
the counts at infrared wavelengths is sub-critical with $\alpha
\sim 0.25$ at a relatively bright magnitude of $K\simeq 18$. This 
contrasts with optical counts which flatten only at very faint limits
(e.g $\alpha\simeq 0.40$ for $V=22-24$ flattening to $\alpha=0.28$
beyond $V=24$ (Smail \etal 1995b)). Moreover, red-infrared colours
(e.g. $I-K$) are more sensitive to redshift than to spectral class in
the redshift range of interest. The opposite is true for say, $B-I$
(see Fig.~\ref{fig-colz}). This leads to a much cleaner and efficient method of
eliminating likely cluster members: we retain both the flat slope and
sufficient numbers required for measurement of the depletion
effect. The major drawback of measuring depletion signals in the
near-infrared until now has been the absence of panoramic infrared
detectors capable of surveying large areas of sky rapidly to the
required depth.  With the commissioning of the Cambridge Infrared
Survey Instrument (CIRSI, Beckett \etal 1998), this becomes a
practicality.

\section{Data}

\subsection{Strategy}

A number of factors enter when considering the merits of undertaking
depletion studies at near-infrared wavelengths.  Foremost, we can
expect the counts to flatten considerably at
cosmologically-significant depths for any passband longward of
1$\mu$m. Whereas classical number-count studies have been almost
exclusively undertaken in the $K$- or $K'$-band (e.g. Moustakas \etal
1997 obtain $\alpha=0.23$ for $18<K<23$; Gardner \etal 1993 obtain
$\alpha=0.26$ for $K>18$), recent counts at $H$ (Yan \etal 1998) show
little change in slope, with $\alpha=0.31\pm 0.02$ for $20<H<24.5$.
Our lensing study will be undertaken using the $H$-band filter.

\begin{figure}
\centerline{\psfig{figure=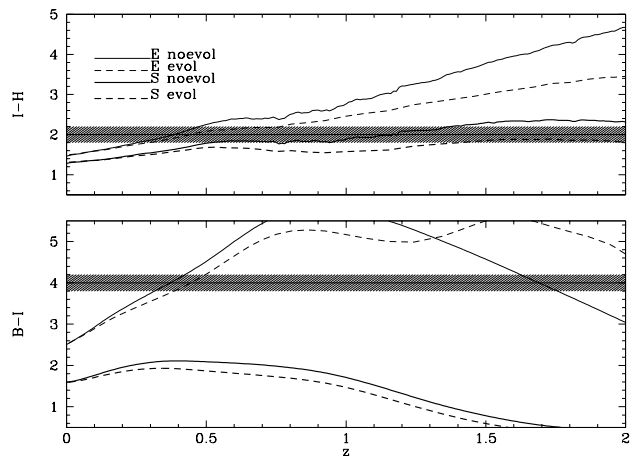,width=0.5\textwidth,angle=270}}
\caption{Dependence of optical-infrared and optical-optical colour on
redshift for early- and late-type galaxies derived from the GISSEL96
spectrophotometric codes (Bruzual \& Charlot, in preparation) with and
without evolution.  The shaded area shows a range of colour when
taking into account typical photometric errors of $\pm 0.2$ for the
measurement of a typical early-type cluster sequence at intermediate
redshift.  Clearly the addition of the infrared magnitude reduces the
dependence of colour on spectral type and allows for a more efficient
selection of a background population.  This is an obvious advantage
over using optical colours alone to distinguish between foreground and
background populations.}
\label{fig-colz}
\end{figure}

Secondly, in terms of colour-selection, the degeneracy between
redshift and spectral class likewise improves dramatically when
infrared magnitudes are added, particularly when account is taken of
likely photometric errors.  Fig.~\ref{fig-colz} illustrates a typical
measurement of the $I-H$ and $B-I$ colour of a cluster early-type
sequence at intermediate redshift.  For the optical-optical colour we
see that the single measurement is not enough to break the degeneracy
between colour and redshift: an object bluer than the cluster sequence
may be a low redshift elliptical or a late-type galaxy at any
redshift.  But when the optical-infrared colour is considered, the
sensitivity to spectral type is greatly reduced, and we see that the
bluer and redder galaxies map more cleanly onto foreground and
background populations.  One can therefore select more red objects
with redshifts greater than that of the cluster. This accurate and
efficient discrimination between the two populations is of great
importance for our depletion study, which depends both on having a
flat number count slope and a sufficient surface number density of
background galaxies.

Our principal goal is to explore the depletion expected in Abell 2219
in the context of earlier studies of this cluster.  This is an X-ray
luminous (Allen \etal 1992) cluster lying at a redshift of 0.22, with
an X-ray temperature of $T_X=11.8$ keV and showing no evidence of a
cooling flow (Allen \& Fabian 1998).  The distribution of the X-ray
gas is elliptical and misaligned with the cD galaxy on small scales.
Smail \etal (1995a) report two systems of giant arcs with undetermined
redshifts: a `red' arc (possible two merging images of a single
background source) and a `blue' arc consisting of three separate
segments.  The red arc is shorter and brighter, and is prominent in
our $H$-band image (Fig.~\ref{fig-arczoom}).

\subsection{Observations}

\begin{table*}
\centering
\begin{minipage}{14cm}
\caption{Summary of observations}
\begin{tabular}{cccccc}
Pointing & $\alpha$ (J2000) & $\delta$ (J2000) & 
total exposure time & seeing\\
\hline
CIRSI $H$-band, cluster centre & 16 40 20.5 & 46 42 44.8  & 5.3 h & 0.9\arcsec \\ 
CIRSI $H$-band, offset fields  & 16 39 50.2 & 46 42 44.8  & 5.5 h & 0.9\arcsec
\\
$I$-band EEV &	16 40 20.5 & 46 42 44.8   & 1.0 h & 0.8\arcsec\\
\end{tabular}
\end{minipage}
\end{table*}

\begin{figure*}
\centerline{\psfig{figure=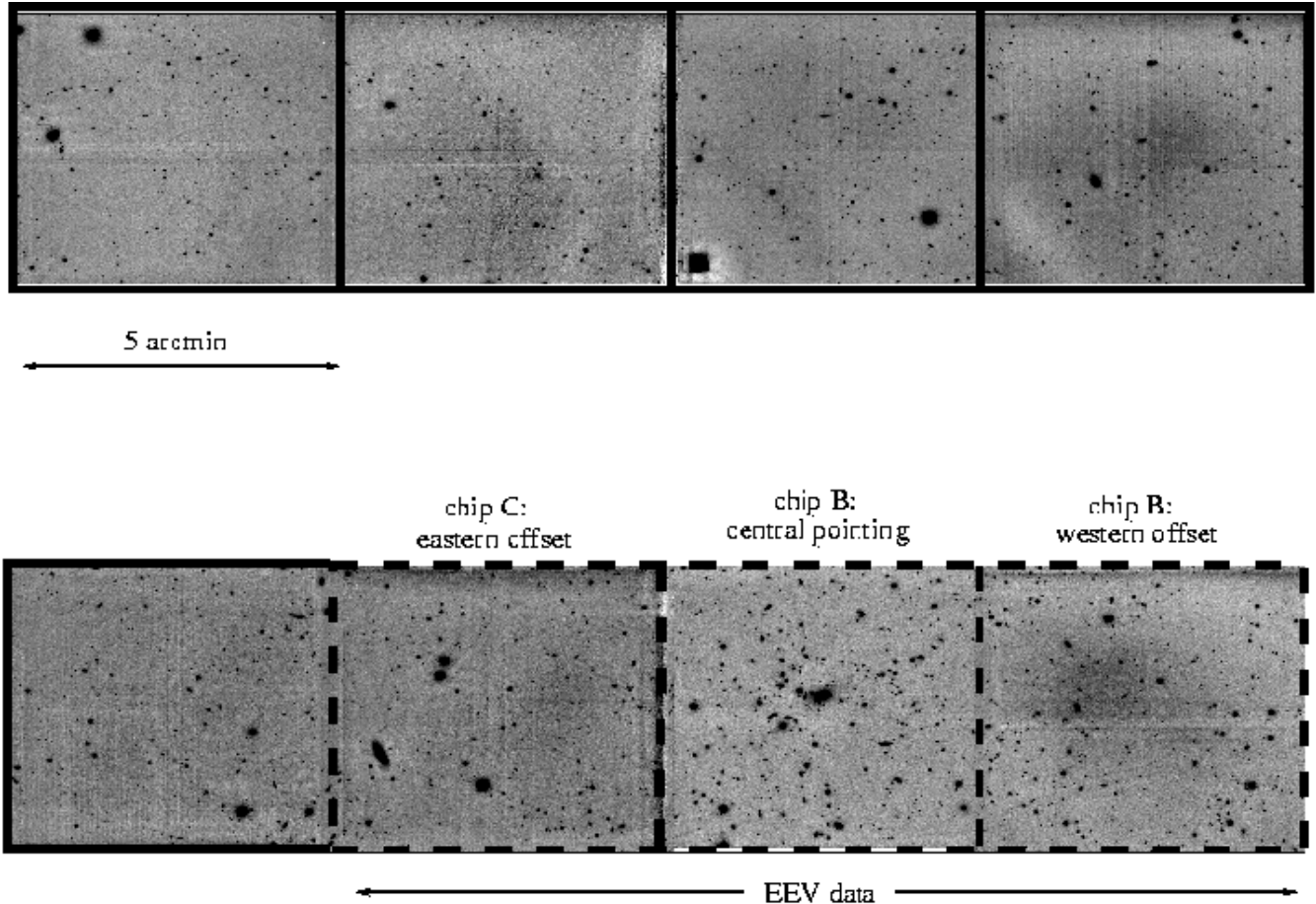,width=1.0\textwidth,angle=0}}
\caption{The field of view of CIRSI after two adjacent pointings.  The
resulting two strips each measure $\sim 20 \times 5$ arcmin.  The
dashed lines indicate the region which overlapped with the optical
data.  Treating the individual CIRSI chips as separate images, the
western offset field, central cluster field, and eastern offset field
(all with overlapping optical data) are labelled.  North is up, east
is to the left.}
\label{fig-fov}
\end{figure*}

The Cambridge Infrared Survey Instrument (CIRSI, Beckett \etal
1998) is a wide-field infrared imager consisting of a mosaic of
four \mbox{1K $\times$ 1K} Rockwell Hawaii HgCdTe detectors.  Each
detector is separated by a gap of approximately one detector width
so that multiple pointings can be easily arranged to produce a
larger contiguously-imaged field.

Table~1 summarizes the observations that form the basis of this
paper. The $H$-band CIRSI observations of \mbox{Abell 2219} were made
at the prime focus of the 4.2-m William Herschel Telescope (WHT); this
offers a plate scale of \mbox{0.32 arcsec pixel$^{-1}$}.  Each CIRSI detector
has a field of view of \mbox{$5.5 \times 5.5$ arcmin} so the mosaic of four
detectors gives an instantaneous total field of view of $11\times 11$
arcmin.

The cluster was observed on the nights 1 to 4 July 1999 using a field
configuration chosen to optimally overlap with existing $B$- and
$I$-band EEV CCD data of plate scale \mbox{0.24 arcsec pixel$^{-1}$}. Two
adjacent CIRSI pointings were arranged to ensure two contiguous strips
of approximately \mbox{$5 \times 20$ arcmin} each, one of which overlaps the
two-pointing \mbox{16 $\times$ 16 arcmin} EEV field (Fig.~\ref{fig-fov}).

The optical CCD data were taken as part of a companion programme and
its reduction is discussed elsewhere (B\'ezecourt et al., 2000).  The
CIRSI observations were taken using short (30 sec) exposures with a 9
point pattern dithered with an offset of 15 arcsec.  Larger offsets of
the order of 50 arcsec were made before starting a new dither
sequence.  Typically 10 exposures were taken at each position, and the
first exposure in each sequence was processed separately due to
instability in the large scale bias level.

Data reduction was done using routines from the CIRSI data reduction
pipeline {\sc cirdr} (Chan \etal 1999, Hoenig \etal 1999) in {\sc
iraf}.\footnote{The Image Reduction and Analysis Facility ({\sc iraf})
is distributed by National Optical Astronomy Observatories, operated
by the Association of Universities for Research in Astronomy, Inc.,
under contract to the National Science Foundation.}  For each dither
sequence, a median sky image was created.  The sky image was then
scaled to and subtracted from each of the individual science
images. The objects were located and a mask for each frame was
constructed using routines from Pat Hall's {\sc phiirs} infrared data
reduction package (Hall, Green \& Cohen 1998).  The original images
were medianed a second time using the object masks to prevent
contamination, and this second-pass sky was again subtracted from each
of the science images.  

The resulting sky-subtracted images were averaged using integer pixel
offsets and a 3-sigma clipping rejection algorithm was applied.  To
construct the final image, the combined dither sequences from each
night's observing were averaged in the same manner. Finally, the
noisier regions at the edge of each mosaic where the overlap of dither
sequences was incomplete were trimmed. Treating each detector as a
separate image, one dither sequence was calibrated using measurements
of a standard star from the UKIRT NIR catalogue (Hawarden
\etal, in preparation). A set of secondary standards within the science image
were then measured and used to calibrate the final mosaiced image,
taking care to consider possible discrepancies in the overlap regions
between the chips.

We used SExtractor 2.0 (Bertin \& Arnouts 1996) to detect sources
and perform isophotal photometry, selecting a detection and analysis
threshold of 1.5$\sigma$ above the sky background.  The $H$-band
images were registered to the $I$-band data and SExtractor was run
in double image mode. This allowed us to detect the objects in the
$I$-band and measure both magnitudes with the same isophote.  The
catalogues were visually inspected and spurious sources removed.

Fig.~\ref{fig-fov} shows the final reduced $H$-band image and indicates the
degree of overlap with the EEV CCD data of B\'ezecourt \etal (2000). 

\subsection{Completeness and Contamination}

\begin{figure}
\centerline{\psfig{figure=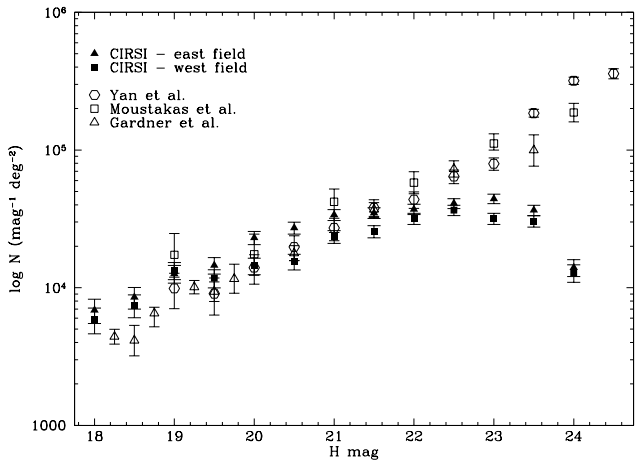,width=0.5\textwidth,angle=270}}
\caption{Comparison of the CIRSI $H$-band number counts with those from
the literature.  Shown are the two CIRSI offset fields (covering 27
arcmin$^2$ each), the Yan \etal (1998) $H$-band survey, and the
Gardner \etal (1993) and Moustakas \etal (1997) $K$-band surveys.  For
comparison purposes the $K$-band points have been shifted by one
magnitude according to the approximation $H-K=1$.  The counts from the
literature have all been corrected for incompleteness at faint
magnitudes.}
\label{fig-counts}
\end{figure}

To understand the magnitude limit of our $H$-band data, we performed
simulations by creating noise-only images: we randomized the offsets
and thus created a misaligned mosaic with the same dimensions and
noise properties as the science images, but with all the objects
removed.  We added artificial galaxies of known magnitudes to these
images using the {\tt artdata} package in IRAF and recovered them
using SExtractor.  We found that for a detection in the $H$-band, the
50\% completeness level (the magnitude at which Sextractor recovered
50\% of the input galaxies) agreed well with the turnover in the
observed number counts at $H=21.7$.  However, for purposes of
measuring colour, we made the detections in the $I$-band and so were
able to push the infrared magnitude limit even deeper.
Fig.~\ref{fig-counts} shows our resulting \mbox{$H$-band} number
counts compared to previous infrared surveys.

In addition to adopting a strictly IR-limited sample, we have also
explored alternative detection strategies noting that we seek the {\it
relative depletion} in the counts as a function of position,
i.e. uniformity of detection is more important than completeness.  We
show in Appendix A that, for the purposes of our depletion
measurements, it is not necessary to have a complete catalogue, so
long as the incompleteness function responsible for the observed
fall-off from the intrinsic power-law distribution is the same for the
offset fields as for the lensed fields.  As we determined the offset
counts simultaneously under identical observing conditions, this holds
true for our sample.  We therefore push the limits of the $H$-band
catalogue past the completeness threshold and adopt a working limit of
$H=24$ for our sample, so that it remains $H$-limited.

As discussed in Appendix~A, it is important to consider not only the
completeness of the sample but also its contamination by spurious
sources arising from noise peaks, especially when the counts are
extended past the completeness limit.  If we define the noise fraction
as the fraction of false objects per magnitude bin, we may proceed
with an incomplete sample only if the noise fraction brighter than our
limiting magnitude is small. To estimate this we used the same pure
noise image discussed above, and measured the magnitudes in 500
randomly placed apertures.  Table~2 shows the resulting distribution
of magnitudes, with only 3.6\% of `detections' falling below the $H=24$
cutoff. This should be considered as an upper limit, since the
detection in our sample is performed in the deeper $I$-band
image. Moreover, contamination, if it is left unaccounted for, will
tend to reduce the lensing signal. We therefore conservatively ignore
this small effect in our analysis.

\begin{table}
\centering
\begin{tabular}{ccr}
magnitude range & number of objects & fraction\\
\hline
$23<H<24$	&  \ 18  &    3.6\% \\
$24<H<30$         &235    & 47.0\%\\
$H=$INDEF         &247    & 49.4\%\\
\end{tabular}
\caption{Noise fraction for the $H$-band image.  Magnitudes were
measured in 500 randomly placed apertures on a pure noise $H$-band
image.  The results show that the number of false objects brighter
than our magnitude limit of $H=24$ is negligible.}
\end{table}
\subsection{Removal of Cluster Members}\label{sec-colmag}

\begin{figure}
\centerline{\psfig{figure=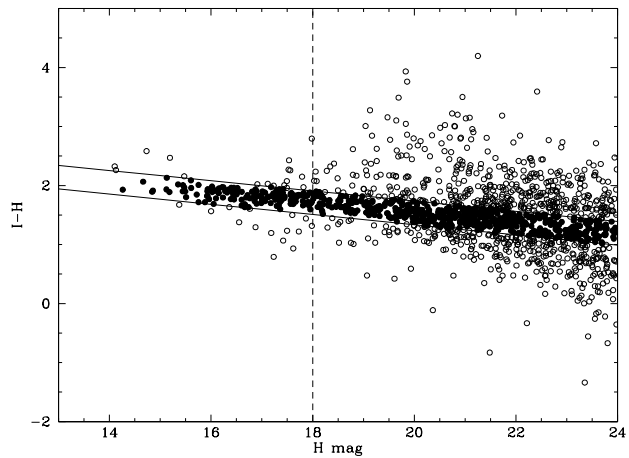,width=0.5\textwidth,angle=270}}
\caption{Colour-magnitude diagram for all objects in the cluster
field.  The cluster galaxies are indicated by filled circles; the
population of objects selected for lensing purposes are those with
colours redder than the cluster sequence and with $18<H<24$ (dashed
line).}
\label{fig-colmag}
\end{figure}

An essential precursor in examining a possible depletion effect in
Abell 2219 is the location of the lensing and unlensed population of
galaxies, the latter of which is, of course, dominated for such a low
redshift system by cluster members. The sequence of cluster galaxies
is clearly visible on the $I-H$ colour magnitude diagram shown in
Fig.~\ref{fig-colmag}. Taking into account known members and
photometric errors, cluster galaxies can be optimally excised
according to the relation
\[
\left| I-H - (3.266 - 0.0865H)\right| < 0.2.
\]

We can divide the remaining population of field galaxies into `red'
and `blue' (unlensed) field populations, adding the further constraint
that $18<H<24$ and noting that the blue population could be composed
of both foreground and blue cluster members.

\begin{figure}
\centerline{\psfig{figure=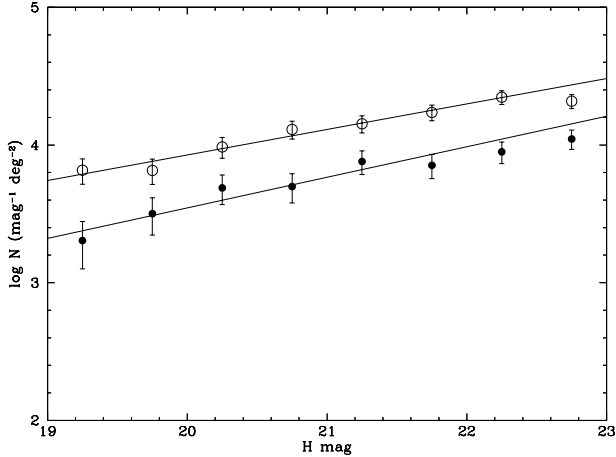,width=0.5\textwidth,angle=270}}
\caption{Number-magnitude relation for non-cluster galaxies in the
offset fields.  Open circles are galaxies redder than the cluster
sequence; filled circles are objects bluer than the cluster sequence.
Note that the slope of the counts for the blue population is not
significantly steeper than that of the red population (as is seen in
the optical, e.g. Taylor \etal (1998)).  Nevertheless we consider only
the red objects for lensing purposes in order to minimize
contamination by foreground galaxies.}
\label{fig-slopes}
\end{figure}

The two CIRSI chips on either side of the central cluster pointing
provide us with flanking fields to characterize the field population.
Combining the catalogues for the east and west flanking fields, we
apply the same colour-selection to determine the properties of the
unlensed background population.  Fig.~\ref{fig-slopes} shows the
resulting number counts for the sub-populations, which are described
by
\[
\log N_{red}=(0.185 \pm 0.017)H + (0.23 \pm 0.35),
\]
and
\[
\log N_{blue}=(0.222 \pm 0.030)H + (-0.89 \pm 0.61).
\]
for the range $19<H<22$ that is well described by a power-law
distribution.  Up to $H=24$, the number densities of the background
populations are $n_0=20.3$ and \mbox{$n_0=10.5$ arcmin$^{-2}$} for the red
and blue populations, respectively.

Finally, to account for obscuration of background galaxies by
foreground cluster members, we created a mask image using the
SExtractor `segmentation' option. This assigns each pixel in the
image a value according to the object that contains it.  We then
used the location of the cluster galaxies (as well as those few
non-cluster objects with $H<18$) selected above to extract those
pixels and form a mask which we used in the subsequent depletion
analysis.

\section{Results}
\begin{figure}
\centerline{\psfig{figure=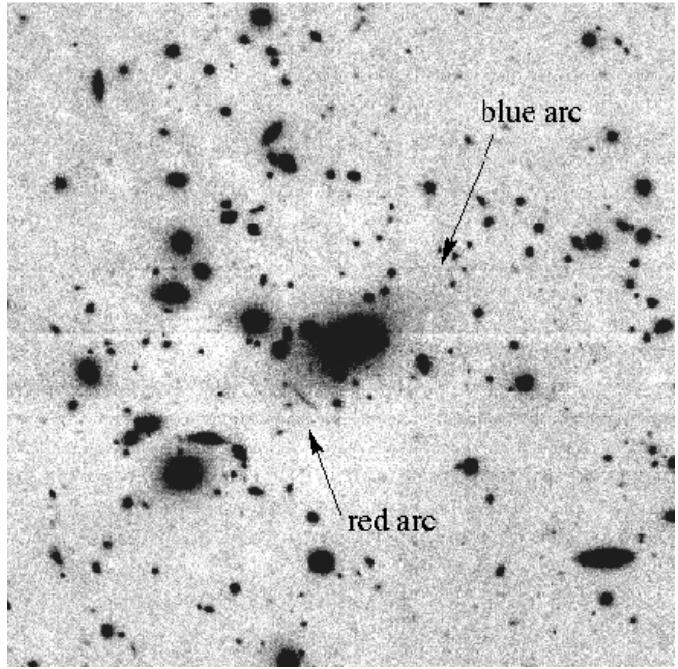,width=0.5\textwidth,angle=0}}
\caption{The central $2.7\times 2.7$ arcmin of the $H$-band CIRSI
image, showing the red arc reported in Smail \etal (1995a).  The
second, bluer arc system is only barely visible in this infrared
image.}
\label{fig-arczoom}
\end{figure}

Our plan will be first to demonstrate the existence of the IR
depletion behind Abell 2219 using the $I-H$ colour diagram to isolate
cluster members and select a background population
($\S$\ref{sec-colmag}).  We will then attempt to quantify the effect
by employing maximum likelihood methods to fit simple, one-parameter
models to the mass distribution for the cluster.  Finally, we will
consider the effect of uncertainties in the adopted mean surface
number density on the maximum likelihood results.

\subsection{Depletion Analysis} 

\begin{figure}
\centerline{\psfig{figure=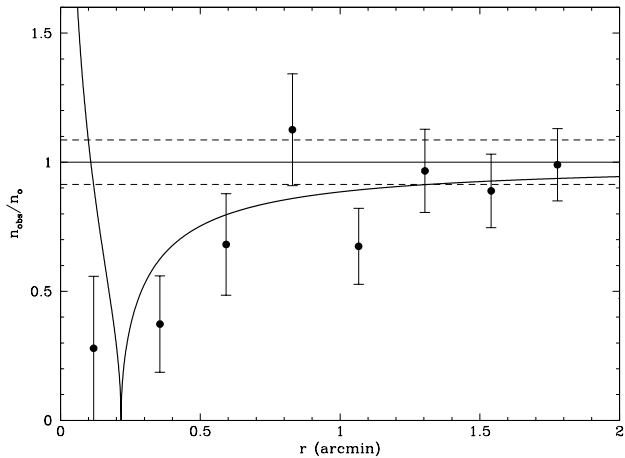,width=0.5\textwidth,angle=270}}
\caption{Ratio of observed number density to unlensed number density
of red galaxies with distance from the cluster centre, using a bin
width of 14 arcsec.  All measurements take into account the area in
each bin obscured by the cluster galaxies.  A depletion of number
counts is clearly seen at small radii.  The dashed lines indicate the
errors in the measured value of $n_0$, the unlensed number density
(the effect of this uncertainty is considered in Section $\S\ref{sec-err}$).
The thick line shows the expected depletion profile induced by a SIS
with $\sigma_v \sim 800$ km s$^{-1}$.}
\label{fig-depletion}
\end{figure}

Fig.~\ref{fig-depletion} shows the radial density profile for field
galaxies selected as discussed in $\S$\ref{sec-colmag}, taking into
account the area masked by the colour-selected cluster galaxies (3.0
arcmin$^2$ in total out of the 27.5 arcmin$^2$ central field).  We
utilize both flanking fields in order to determine the unlensed
surface number density of the red population to be \mbox{$n_0$=20.3
arcmin$^{-2}$}.  With no colour selection this density rises to
$n_T$=43.2 arcmin$^{-2}$ which gives some indication of the likelihood
that non-cluster galaxies might be discarded via this colour cut.

Fig.~\ref{fig-depletion} gives a clear detection of the depletion
signal at the 3$\sigma$ level within a diameter of $\sim$80 arcsec
($\sim$350 kpc for $h=0.5$, $q_o=0.5$).  However, its absolute
significance depends critically on the adopted value of $n_0$.  We
estimate the uncertainty in our measurement of $n_0$ (indicated by the
dashed lines in Fig.~\ref{fig-depletion}) using the dispersion in
counts in the full $H$-band dataset shown in Fig.~\ref{fig-fov}.  The
effects of this uncertainty are discussed further in
$\S\ref{sec-err}$.

To investigate the magnitude and significance of this possible
depletion, we adopted a maximum likelihood approach based on that
developed by Schneider, King \& Erben (2000; hereafter SKE). In this
formulation, which avoids the loss of information induced by the
radial binning (as in Fig.~\ref{fig-depletion}), the log-likelihood
equation takes the form:

\begin{equation}\label{eqn-l}
l = -n_0 \int d^{2}\theta[\mu({\mathbf \theta)}]^{2.5\alpha -
1} + (2.5\alpha - 1)\sum_{i=1}^{N}\ln\mu({\mathbf \theta}_i),
\end{equation}
where $n_0$ is the unlensed number density of background galaxies
(arcmin$^{-2}$), $\mu$ is the magnification, ${\mathbf \theta}_{i}$ is
the position vector of the $i$th galaxy in the field with respect to
the cluster centre, $N$ is the total number of galaxies observed, and
\mbox{$\alpha=d\log{N}/dm$} is the intrinsic logarithmic slope of the
number counts.  The first term in the log-likelihood function
addresses the probability of finding $N$ galaxies in the field of view
given the lens model $\mu(\theta)$ and the population parameters
$\alpha$ and $n_0$, while the second term concerns the probability of
finding each galaxy $i$ at position $\theta_i$.  By maximizing $l$ we
find the most likely parameter(s) for a given lens model. In the
subsequent sections, we examine two single-parameter mass models.

The derivation of this expression is presented in
Appendix~A. Importantly, this appendix also generalizes the likelihood
method of SKE to include the effect of incompleteness. We showed
that, as long as the intrinsic unlensed counts follow a power law,
incompleteness is very simple to account for: one simply uses the
observed (incomplete) unlensed density $\tilde{n}_{0}$ in place of the
intrinsic (complete) unlensed density $n_{0}$ in the likelihood
function.  This allows us to include, in our sample, galaxies which
are fainter than the completeness limit and therefore of improving our
signal-to-noise, without fear of introducing a bias.

\subsubsection{Model 1: Singular isothermal sphere}

First, we model the cluster according to a singular isothermal sphere (SIS)  
parametrized with the Einstein radius, $\theta_{\rm E} $:
\begin{equation}\label{eqn-musis}
\mu_{\rm SIS}(\theta)=\left| \frac{1}{1-\theta_{\rm E}/\theta} \right|.
\end{equation}

\begin{figure}
\centerline{\psfig{figure=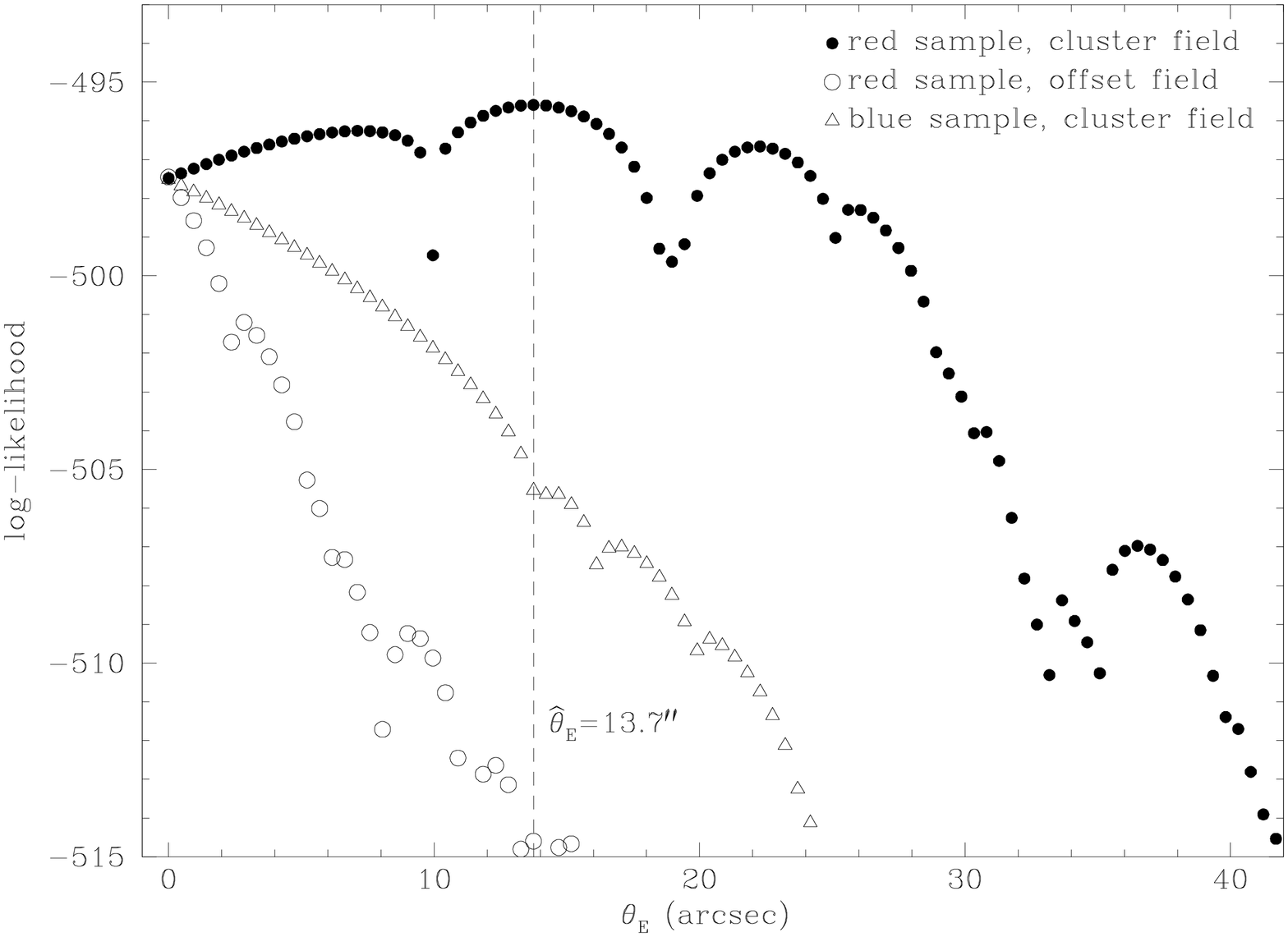,width=0.5\textwidth,angle=270}}
\caption{Maximum likelihood analysis for depletion of the red
background galaxy sample (solid circles).  The model corresponding to
the peak is a SIS parametrized by the Einstein radius
\mbox{$\hat{\theta}_E=13.7$ arcsec}.  By contrast, the same analysis
for the red galaxy sample in one of the offset fields (open circles)
peaks at \mbox{$\hat{\theta}_E=0$ arcsec} and shows no evidence of
depletion, as does the sample of blue (foreground) galaxies in the
cluster field (open triangles).  The curves have been vertically
shifted to the same zeropoint for clarity.}
\label{fig-mlplot}
\end{figure}

The filled circles in Fig.~\ref{fig-mlplot} show the resulting
log-likelihood curve.  The sharp dips are a result of the contribution
of the second term of Equation~\ref{eqn-l} (concerning the galaxy
positions) to the log-likelihood function.  As the likelihood is
calculated for increasing values of the input $\theta_E$, a galaxy may
happen to lie on the critical line ($\theta_i=\theta_E$). Since the
magnification (and hence the depletion) is formally infinite at this
radius, the probability of finding a galaxy vanishes, and the method
rejects this particular model.  This results from the fact that we
have effectively modelled the galaxies as point sources. In practice,
galaxies are extended objects and therefore are not subject to
infinite magnifications, but instead are stretched into giant arcs on
the critical line. In our analysis, we simply identified the peak as
the absolute maximum, without any interpolation. As we will see below,
this does not introduce any observable bias in the determination of
the model parameters.

The peak of the likelihood function is reached at 
\mbox{$\hat{\theta}_{\rm E}=13.7$ arcsec}, which is consistent with the 
location of the red giant arc located 13 arcsec from the cluster centre
(shown in Fig.~\ref{fig-arczoom}). 

We perform a test of the depletion by applying the same
maximum-likelihood test on one of the offset fields.  Using the same
colour selection criteria to define a population of red objects we
find no evidence for any depletion effect.  Furthermore, we also test
the population of `foreground' galaxies (those with colours bluer than
the cluster sequence) in the cluster field, and again find as expected
no evidence for depletion.  The log-likelihood functions for these two
samples, both peaking at $\theta_E=0$ arcsec, are also shown as open
symbols in Fig.~\ref{fig-mlplot}.

\subsubsection{Model 2:  The NFW profile}

An alternative model to the SIS is the universal density profile for
dark matter halos proposed by Navarro, Frenk \& White (1997, 1996,
1995).  Given the uncertainties in the data it is unlikely that we
could use this technique to distinguish between two models.  It is
nevertheless of interest to demonstrate how the method could be
applied to a second model.

The NFW density profile follows
\begin{equation}\label{eqn-rho}
\rho(r) = \frac{\delta_c \rho_c}{(r/r_s)(1+r/r_s)^2},
\end{equation}
where $\rho_c$ is the critical density at the redshift of the lens.
The two parameters of the model are contained in the scale radius
$r_s$ and the concentration parameter $c$, so that the characteristic
overdensity is
\begin{equation}\label{eqn-dc}
\delta_c = \frac{200}{3} \frac{c^3}{\ln(1+c)-c/(1+c).}
\end{equation}

In contrast to the SIS, this profile  flattens towards the core and is
capable  of producing radial as  well  as tangential arcs  (Bartelmann
1996).  For  a   projected radial   distance  $R$,  we  can  define  a
dimensionless distance, \mbox{$x=R/r_s$}.   Wright  \& Brainerd (1999)
give the formulation for the radial dependence  of the surface density
$\Sigma(x)$ and the shear $\gamma(x)$.  The convergence is then simply
$\kappa(x)=\Sigma(x)/\Sigma_{\rm   crit}$,  where the critical surface
number mass density
\begin{equation}\label{eqn-sigma}
\Sigma_c = \frac{v_c^2}{4\pi G} \frac{D_s}{D_d D_{ds}}
\end{equation}
depends on the angular diameter distances $D_s,D_d,D_{ds}$ from observer
to source, observer to lens, and source to lens, respectively, and the
speed of light $v_c$.  The magnification is then simply
\begin{equation}
\mu_{\rm NFW}(x)= \left| \frac{1}{(1-\kappa(x))^2 - \gamma(x)^2)}\right|,
\end{equation}
(e.g. Schneider, Ehlers and Falco 1992).

Using $\mu_{\rm NFW}$ in Equation~\ref{eqn-l}, we perform the maximum
likelihood analysis for an NFW profile on Abell 2219.  We reduce the
model to one parameter by assuming a reasonable value for the
concentration parameter, namely $c=6$, and fitting for $r_s$.  We find
a maximum at $r_s$ corresponding to an angular length of 52 arcsec.
At the redshift of the cluster this gives a scale length of
125$h^{-1}$ kpc (where $H_0=100h$ km s$^{-1}$ Mpc$^{-1}$), which
compares well to the $r_s \sim 250 h^{-1}$ predicted from numerical
simulations of CDM models (Bartelmann 1996).

\begin{figure}
\centerline{\psfig{figure=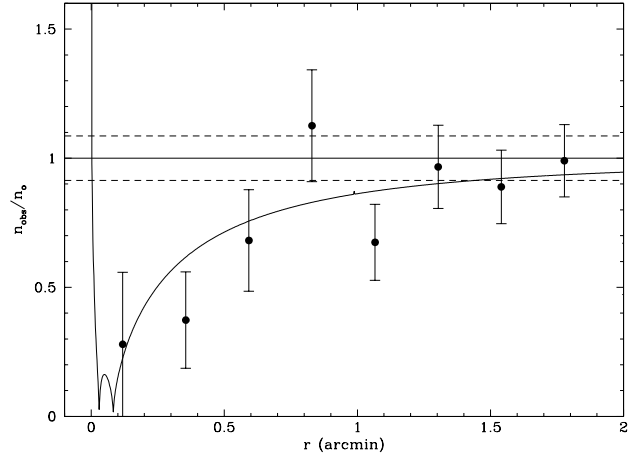,width=0.5\textwidth,angle=270}}
\caption{The radial depletion curve for a NFW cluster with $r_s = 125 h^{-1}$
kpc and $c=6$.  The inner dip corresponds to the radial arc at $x_r=2$
arcsec while the outer dip corresponds to the tangential arc at
$x_t=5$ arcsec, for a source population with $\alpha=0.185$.
Overlayed are the points the Abell 2219 depletion curve, as in
Fig.~\ref{fig-depletion}.}
\label{fig-nfw}
\end{figure}

Fig.~\ref{fig-nfw} shows the form of an NFW depletion curve
resulting from the substitution of $\mu_{\rm NFW}$ in
Equation~\ref{eqn-N}, for such a cluster with $c=6$ and $r_s=125
h^{-1}$ kpc .  The observed depletion curve of Abell 2219 is
overplotted for comparison, and we see that the data are also
consistent with this model.  However, the data do not allow us to
distinguish between the NFW and SIS models.  In the further analysis
we choose to continue with the SIS for simplicity.

\subsection{Simulations}\label{sec-sim}

To further test the robustness of the depletion effect we have
simulated a population of galaxies with the same properties as our
putative background population.  In our observed sample the intrinsic
power law form of the background number counts is modified by the
incompleteness function.  As shown in Appendix~A, it is however
equivalent to use a renormalized power law for the counts in the
simulations.  Specifically, we keep the same observed slope,
$\alpha=0.185$ but lower the normalization to maintain the same
observed unlensed number density $\tilde{n}_o=20.3$ arcmin$^{-2}$.

\begin{figure}
\centerline{\psfig{figure=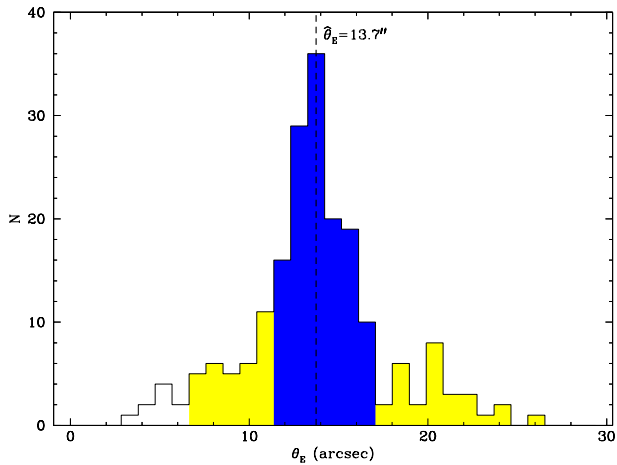,width=0.5\textwidth,angle=270}}
\caption{Distribution of the recovered $\theta_{\rm E}$ from 200
realizations of simulated lensing by an isothermal sphere, using the
observed background population parameters (\mbox{$n_0=20.3$
arcmin$^{-2}$}, \mbox{$\alpha=0.185$}).  The darker shaded region
indicates the resulting 66\% confidence interval in the range
\mbox{$11.1<{\theta}_E<17.06$} arcsec, centred around the input value
of \mbox{$\hat{\theta}_{\rm E} = 13.7$ arcsec}. The lighter shaded
region indicates the 95\% confidence interval in the range
\mbox{$6.8<{\theta}_E<26.5$} arcsec.}
\label{fig-mldist}
\end{figure}

Scattering the simulated galaxies randomly across the field of view,
we then adjusted their positions and magnitudes as if they were lensed
by a singular isothermal sphere with $\theta_{\rm E} = 13.7$
arcsec. We created a catalogue of those galaxies whose lensed
magnitudes were brighter than our limiting magnitude and whose lensed
positions were not obscured by the Abell 2219 cluster mask and applied
the same maximum likelihood technique.  Fig.~\ref{fig-mldist} shows
the distribution of the resulting best-fitting models for 200
realizations.  We see that the distribution of recovered Einstein
radii is centred around the input value of 13.7 arcsec, and deduce the
95\% confidence interval to span the range
\mbox{$6.8<\hat{\theta}_E<26.5$ arcsec}.

To translate the estimates of \mbox{$\theta_{\rm E}$} into 
an statement regarding the cluster mass, we use the relation for a SIS:
\begin{equation}\label{eqn-te}
\theta_{\rm E} = 4\pi \frac{\sigma^2_v}{c^2}\frac{D_{\rm ds}}{D_{\rm s}},
\end{equation}
where $\sigma^2_v$ is the velocity dispersion of the lens, $D_{\rm
ds}$ and $D_{\rm s}$ are the angular-diameter distances from lens to
source and from observer to source, respectively.  

Modelling the background $N(z)$ from the $K$-band surveys of Cowie
\etal (1996), we find the median redshift for the background is
$\left< z \right>=1.0$ for $H<22$, increasing to $\left< z
\right>=1.3$ at $H<24$.  Given $z_{\rm lens}=0.22$ and assuming a
cosmology of \{\mbox{$H_0=50 $ km s $^{-1}$ Mpc $^{-1}$}, $q_o=0.5$\}
we obtain $\sigma_v = 842^{+112}_{-141}$ km s$^{-1}$ for $\left< z
\right>=1$ and $\sigma_v = 814^{+109}_{-136}$ km $s^{-1}$ for $\left< z
\right>=1.3$.  The quoted errors refer to the 66\% confidence limits
derived from the above simulations.  Clearly the result is more
sensitive to the uncertainties in the measurement of $\theta_E$ than
the uncertainties in the background redshift distribution.  Therefore,
we consider the background to lie on a single sheet at the median
redshift, and placing the galaxies at $z=1.3$ and adding the errors in
quadrature we obtain our final result of $\sigma_v=814^{+112}_{-139}$
km s$^{-1}$ (66\% confidence level).  Our SIS model is in good
agreement with the result from Smail \etal (1995a), who find $\sigma_v
\sim 930$ km s$^{-1}$ using strong lensing constraints.  Note that the
precision of the strong lensing constraint would be tighted if the
redshift of the red arc were known.  Our results are also consistent
with the optical weak shear analysis of B\'ezecourt et al. (2000), who
find $\sigma_v = 1075\pm 100$ km s$^{-1}$ from an optical weak shear
analysis.  In principle, by adding one additional parameter to our
model we could also constrain the amplitude of a mass-sheet, and
therefore break the degeneracy inherent in reconstructions involving
shear only.  However, fitting an extra parameter is not warranted by
the data, especially after taking into account the uncertainty in
$n_0$ (see \S4.3 below).

\begin{figure}
\centerline{\psfig{figure=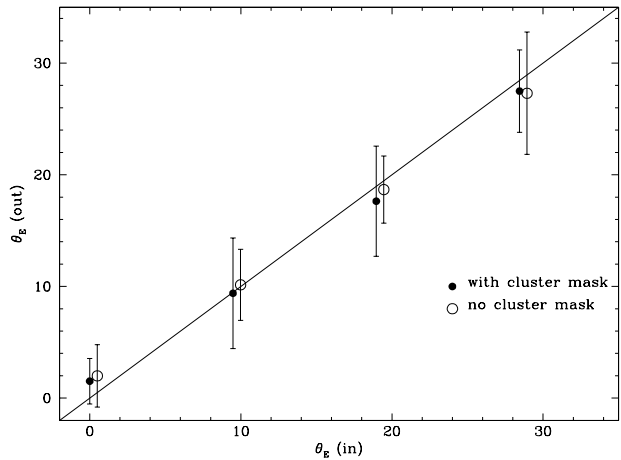,width=0.5\textwidth,angle=270}}
\caption{Simulations showing recovered $\theta_E$ vs input $\theta_E$, with
and without a cluster mask (the open circles are offset slightly in
the $+x$ direction for clarity).  The cluster mask, which
obscures about 10\% of the field including much of the central region,
has no significant effect on the recovered $\theta_E$ (aside from
increasing the errors), even given the low numbers of background
galaxies.}
\label{fig-sim}
\end{figure}

The above simulations only enable us to deduce the error on $\theta_E$
assuming our initial estimate of the model parameters is correct.  To
test the method across a wider range of input lenses, we have
simulated lenses with input parameters \mbox{$\theta_{\rm E} = 0, 10,
20,$ and 30 arcsec}, and performed the maximum likelihood analysis 25
times in each case.  Fig.~\ref{fig-sim} shows the recovered
$\hat{\theta}_{\rm E}$ vs the input $\theta_{\rm E}$, both with and
without the cluster mask.  We see that neither the dips in the
likelihood function, nor the masked area introduce any significant
bias into the method. In addition, we find that obtaining a
best-fitting value of $13.7$ in the absence of lensing, i.e. a false
detection, is unlikely, and we reject this null hypothesis at the
6$\sigma$ level. This figure provides a measure of the significance of
our detection of the lensing signal if the uncertainty in $n_0$ is
ignored and the assumed model is correct.
 
\subsection{Effect of uncertainty in number counts}\label{sec-err}

SKE demonstrate that, to first order, most of the magnification
information is provided by the unlensed number density $n_0$.  Without
adequate knowledge of this normalization, the mass-sheet degeneracy
cannot be broken. Prior knowledge of the uncertainty in $n_0$ can be
included in the maximum-likelihood analysis.  Assuming that the true
$n_0$ follows a Gaussian distribution with mean $\bar{n}_o$ and
dispersion $\sigma_n=\eta\bar{n}_o$ then, as shown in SKE, the
log-likelihood function becomes:

\begin{equation}\label{eqn-lerr}
l(\eta) = -n_0I+ (\beta-1)\sum_{i=1}^{N}\ln\mu({\mathbf \theta}_i) +
N\ln n_0 - \left( \frac{(n_0-\bar{n}_o)^2}{2\left( \eta\bar{n}_o\right)^2}\right),
\end{equation}
where $I=\int d^{2}\theta[\mu({\mathbf \theta)}]^{\beta -1}$ and
$\beta=2.5\alpha$.  This can be maximized with respect to $n_0$,
yielding
\begin{equation}\label{eqn-lerr2}
\frac{n_0}{\bar{n}_o} = \frac{1}{2}\left(1-\eta^2\bar{n}_oI\right) +
\sqrt{\frac{1}{4}\left(1-\eta^2\bar{n}_oI\right)+ \eta^2N}.
\end{equation}
Substituting this value for $n_0$ into Equation~\ref{eqn-lerr} yields
the new log-likelihood function.

To determine the value of $\eta$ for our own sample, we turn to the
entire $H$-band dataset pictured in Fig.~\ref{fig-fov}.  While the sample used in
the depletion analysis is a colour-selected sample that incorporates
the $I$-band data, the area in common with the optical image allowed
for only the two offset fields discussed previously.  However, we
attempt to estimate the uncertainty in $n_0$ by sampling the counts in
the entire $H$-band region.  

We divided the two $H$-band strips into four single-chip images each,
discarding the chip containing the cluster and the two images taken
with chip 3 (which was contaminated by many hot pixels, leading to a
high number of false detections).  This left five offset images of
equal area.  We constructed $H$-limited catalogues on the central
\mbox{$4.3\times4.3$ arcmin} region of each image using SExtractor and
examined the dispersion in the total $H$-band number counts in the
range $18<H<24$ for each chip.  The resulting distribution of raw
counts, listed in Table~3, has mean $\bar{N}=467$ and standard
deviation $\sigma=40$, hence the fractional error is $\eta_{\rm
H}=0.086$.  If the fraction of galaxies discarded during the colour
selection to form the `red' sample of background galaxies is constant
from field to field (from the two multi-colour offset fields it
appears to be $\sim 50\%$), then this error will also apply to the
subpopulation.  

\begin{table}
\centering
\begin{tabular}{cc}
field & N($18<H<24$)\\
\hline
1	& 	468\\
2	&	516\\
3	&	483\\
4	&	476\\
5	&	394\\
\end{tabular}
\caption{$H$-band counts in 18.2 arcmin$^2$ regions in five of the
seven offset fields, yielding a fractional uncertainty of 8.6\%.}
\end{table}

We note that this is higher than the Poisson error for these number
counts.  However, if we consider the angular correlation function
$\omega(\theta)$ of $K$-band galaxies to \mbox{$K=21.5$} presented in
Carlberg \etal (1997) and extrapolate to \mbox{$\theta=2.5$} arcmin
then \mbox{$\omega(\theta) \sim 0.01$}.  This yields a rough estimate of
\mbox{$\delta N/N \sim \sqrt{\omega(2.5^{\prime})} \sim 0.1$} which is in agreement with our measurement of
the dispersion.  Our uncertainty in the unlensed density is therefore
dominated by clustering.

\begin{figure}
\centerline{\psfig{figure=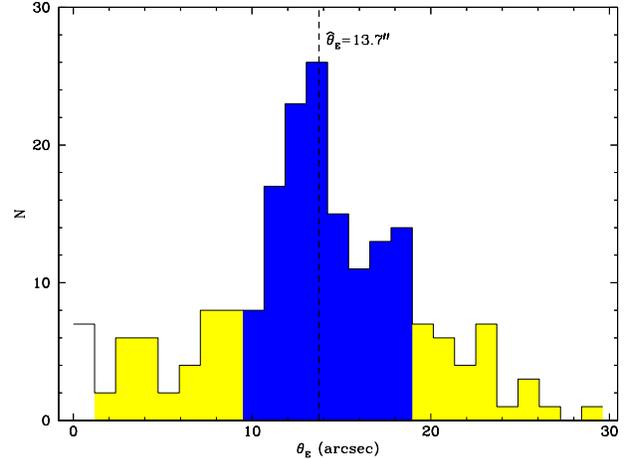,width=0.5\textwidth,angle=270}}
\caption{The effect of uncertainty in the background number counts
from simulations.  Shown is the distribution of the recovered
$\theta_E$ for 200 realizations, with the true $n_0$ drawn from a
Gaussian distribution with mean 20.3 arcmin$^{-2}$ and fractional
error $\eta=8.6$\%.  The lensed populations were analysed as in
Equations~\ref{eqn-lerr}~and~\ref{eqn-lerr2} using the estimated
$\bar{n}_o$=20.3 arcmin$^{-2}$.  The 66\% and 95\% confidence limits
are indicated by the dark (\mbox{$8.9<{\theta}_E<18.9$}) and light
(\mbox{$0.9<{\theta}_E<29.6$}) shaded regions.  A comparison with
Fig.~\ref{fig-mldist} shows that incorporating a significant
uncertainty in the analysis increases the error bars on the
measurement of $\theta_E$.}
\label{fig-errsim}
\end{figure}

We again turn to simulations to derive the error bars on our
measurement.  As in $\S\ref{sec-sim}$ we simulate a population that is
lensed by a SIS with $\theta_E=13.7$ arcsec, but in this case we draw
the true value of $n_0$ from a Gaussian distribution with mean 20.3
arcmin$^2$ and dispersion $\sigma_n=\eta n_0$, where
\mbox{$\eta=0.086$} as derived above.  Using the log-likelihood
function in Equation~\ref{eqn-lerr} and a fixed estimate
\mbox{$n_0=20.3$}, we show the resulting distribution of recovered
$\theta_E$ for 200 realizations in Fig.~\ref{fig-errsim}.  We see, by
comparison with Fig.~\ref{fig-mldist} (the case where $n_0$ was known
precisely), that incorporating an error of \mbox{$\eta=0.086$}
produces a much broader distribution, with the result of greatly
increasing the error bars on our measurement.  Applying
Equation~\ref{eqn-lerr} to the real data, we find
$\hat{\theta}_E=1.4^{+5.2}_{-1.4}$ arcsec (66\% confidence) or
$\hat{\theta}_E=1.4^{+15.9}_{-1.4}$ (95\% confidence).  While the
estimate of $\theta_E$ itself is greatly reduced, the error bars make
it compatible with the result of $\S\ref{sec-sim}$ within a 2$\sigma$
range.

This section illustrates the principal weakness of the depletion
method as pointed out in SKE, namely the vital importance of
accurately measuring $n_0$.  With a $\delta N/N \sim 0.1$ which is not
unreasonable given clustering on these scales, the depletion effect is
made more difficult to distinguish from variations in the background
counts that are not due to lensing.  However, this effect could be
countered by selecting similar clusters (e.g. by their X-ray
temperatures) and stacking the depletion signal accordingly, thus
calibrating the cluster $T_X$-mass relation and obtaining an average
cluster mass profile.  Choosing a sample of 10 clusters would
simultaneously increase the effective background surface number
density by a factor of 10 while reducing the fractional error in $n_0$
by a factor of $\sqrt{10}$.  Simulations show that for 10 clusters
similar to Abell~2219, the 95\% confidence region on $\theta_E=13.7$
arcsec would then shrink from \mbox{$0.9 < \theta_E < 29.6$ arcsec}
(as quoted above) to only
\mbox{$7.7 < \theta_E < 16.1$ arcsec}; a feasible project for upcoming 
IR survey telescopes.  Note that in practice scatter in the cluster
properties would increase this range somewhat.

\section{Conclusions}

We present a study of the depletion effect around Abell 2219, the
first done in the infrared, using the newly-available panoramic IR
camera (CIRSI). We show (see Appendix A) that the sample can be
effectively exploited beyond the completeness limit, as long as the
lensed field and the background field obey the same incompleteness
functions and have minimal contamination by false objects. This allows
us to detect a clear dip in the radial number density profile of
background galaxies at small distances from the cluster centre.  The
optical-infrared colours enable us to select a population of red
background galaxies with a flat number-count slope in order to
optimise the lensing signal and reduce foreground-background
confusion.

For a population of red background galaxies with an extremely flat
slope ($\alpha=0.185$), we employ maximum likelihood methods and a SIS
model to derive an estimate of the Einstein radius
$\theta_E=13.7^{+3.9}_{-4.2}$ arcsec (66\% confidence limit when
uncertainties is $n_0$ are ignored), resulting velocity dispersion
$\sigma_v=814^{+112}_{-139}$ km s$^{-1}$.  These values are consistent
with the location of the redder of the two giant arcs and the estimate
$\sigma_v \sim 930$ km s$^{-1}$ of Smail \etal (1995a).

We examine the uncertainty in the number counts, and derive a
fractional error of 8.6\% on the normalisation of the background
number density (consistent with clustering on these scales).  When
this error is incorporated into the maximum-likelihood analysis the
error bars become too large to make a precise statement about the
magnitude of the lensing (although our previous measurement is not
ruled out).  This demonstrates the crucial importance of the
background density $n_0$ for an accurate depletion analysis as discussed
in Schneider, King \& Erben (2000).

Finally, while we cannot at present use our current data to
distinguish between alternative models for the cluster mass
distribution, we demonstrate the application of the method to the data
using a second mass model.  We show the typical depletion curve
induced by a cluster with an NFW density profile, and using this
profile as a second estimate of the cluster mass we derive a scale
length of 125$h^{-1}$ kpc for the cluster.

The wide field IR coverage available using CIRSI is an ideal match to
the current generation of wide-field optical imagers.  The availability of
multicolour data taken simultaneously but at large radii from the
cluster centre provides offset information about the field population
and eliminates the need to normalise the depletion curve to the values
at the edge of the chip, and the infrared sample provides a cleaner
discriminant between foreground and background and a flat number-count
slope.

The depletion method is an elegant and relatively simple way to derive
mass estimates of clusters of galaxies.  Deeper IR surveys will
provide more accurate knowledge of $\omega(\theta)$ on arcminute
scales, allowing us to better understand the degree to which
clustering affects the method.  Future studies of clusters with
wide-field optical-infrared data (e.g. with the VISTA telescope,
\mbox{\tt http://www-star.qmw.ac.uk/$\sim$jpe/vista}) covering a wide
wavelength range could provide more accurately selected background
populations via photometric redshifts and allow us to add another
sample of independent mass profiles to be compared with those derived
from velocity dispersions, X-ray measurements, and strong and weak
lensing.  The problem posed by the uncertainty in the background
number counts can be overcome by selecting similar clusters (e.g. by
their X-ray temperatures) and stacking the depletion signal
accordingly to obtain an average cluster mass profile; a feasible
project for future IR survey telescopes.

\section*{Acknowledgments}
We thank Peter Schneider, Lindsay King, Andy Taylor and Konrad Kuijken
for useful discussions, and Felipe Menanteau for assistance with the
GISSEL96 models.  The Cambridge Infrared Survey Instrument is
available thanks to the generous support of Raymond and Beverly
Sackler.  MEG wishes to acknowledge the support of the Canadian
Cambridge Trust and the Worshipful Company of Scientific Instrument
Makers. AR was supported by the European TMR Lensing network and by a
Wolfson College Fellowship.  This research has been conducted under
the auspices of the European TMR network ``Gravitational Lensing: New
Constraints on Cosmology and the Distribution of Dark Matter'', made
possible via generous financial support from the European Commission
\mbox{({\tt http://www.ast.cam.ac.uk/IoA/lensnet}).}

\appendix
\section{Effect of Incompleteness}
The likelihood function for the lensing depletion has been derived by
SKE, for the case of a complete sample of background galaxies. Here
we generalize their results to include the effect of incompleteness in
the sample. 

Let the unlensed differential counts of the background galaxies be
denoted by $\frac{dn_{0}(S)}{dS}$, where $n_{0}$ is the number of
galaxies per unit solid angle, and $S$ is the flux. The lensed counts
are given by
\begin{equation} 
\label{eq:dnds_lens}
\frac{dn(S)}{dS} = \frac{1}{\mu^{2}}
\frac{dn_{0}(\frac{S}{\mu})}{dS}, 
\end{equation} 
where $\mu$ is the magnification. The observed (lensed) counts, 
in the presence of incompleteness, can be written as
\begin{equation}
\frac{d\tilde{n}(S)}{dS} = \frac{dn(S)}{dS} \phi(S),
\end{equation}
where $\phi(S)$ is the completeness fraction for a given flux
$S$. Typically, $\phi=1$ for large fluxes, and $\phi<1$ for small
fluxes. In the following, we will continue to use the subscript $_{0}$
to denote unlensed counts, and a tilde to denote the observed
counts. In practice, it is convenient to consider the total counts
down to a detection limit $S$
\begin{equation}
n(>S) = \int_{S}^{\infty} dS' \frac{dn(S')}{dS}.
\end{equation}

In our case (see Fig.~\ref{fig-slopes}) and for many applications, the
intrinsic unlensed counts are well described by a power law of the
form
\begin{equation}
\frac{dn_{0}}{dS} \propto S^{-\beta-1},~~~n_{0}(>S) \propto S^{-\beta},
\end{equation}
where $\beta$ is the flux slope parameter, which is related to the
magnitude slope parameter $\alpha \equiv d(\log n_{0})/dm$
(Equation~\ref{eq:alpha}) by $\beta=2.5\alpha$. In this case,
Equation~(\ref{eq:dnds_lens}) takes the simple form
\begin{equation}
n(>S) \left/ n_{0}(>S) \right. = 
  \frac{dn}{dS} \left/ \frac{dn_{0}}{dS} \right.= \mu^{\beta-1}.
\end{equation}
In this case, the observed lensed counts are given by
\begin{equation}
\label{eq:dnds_obs}
\frac{d \tilde{n}}{dS} = \phi \mu^{\beta-1} \frac{dn_{0}}{dS}.
\end{equation}

As a result, the mean integrated lensed counts $\langle \widetilde{N}
\rangle$ observed in the field are given by
\begin{equation} 
\langle \widetilde{N} \rangle = 
\int d^{2}\theta \int_{S}^{\infty} dS \frac{d\tilde{n}}{dS} =
\tilde{n}_{0} \int
d^{2}\theta \left[ \mu(\mathbf{\theta}) \right]^{\beta-1}, 
\end{equation} 
where $\tilde{n}_{0} = \int_{S}^{\infty} dS' \phi \frac{dn_{0}}{dS}$
is the observed surface number density of unlensed galaxies.  

The probability of finding the $i^{\rm th}$ galaxies at position
${\mathbf \theta}_{i}$ is $p_{i} \propto \tilde{n}({\mathbf
\theta}_{i}) = \tilde{n}_{0} [\mu({\mathbf
\theta}_{i})]^{\beta-1}$. Thus, the likelihood of finding
$N$ galaxies in our incomplete survey at positions
${\mathbf \theta}_{i}$ is given by
\begin{equation}
{\mathcal L} = P(N;\langle \widetilde{N} \rangle)
  \prod_{i=1}^{N} \frac{[\mu({\mathbf
\theta}_{i})]^{\beta-1}}{ \int d^{2}\theta [\mu({\mathbf
\theta}_{i})]^{\beta-1}},
\end{equation} 
where $P(N;\langle N \rangle)$ stands for the Poisson distribution
with mean $\langle N \rangle$.  The log likelihood function is thus
\begin{equation}
l = - \tilde{n}_{0} \int d^{2}\theta [\mu({\mathbf
\theta}_{i})]^{\beta-1} + (\beta -1) \sum_{i=1}^{N}
\ln{\mu({\mathbf \theta}_{i})},
\end{equation}
where constant terms have been dropped. 

These two expressions are identical to the corresponding expressions
in SKE, except that incomplete counts $\langle \widetilde{N}
\rangle$ and $\tilde{n}_{0}$ replace the complete counts $\langle N
\rangle$ and $n_{0}$, respectively.  Therefore, the effect of 
incompleteness is very simple to take into account. (We will drop the
tilde in the rest of the text). In fact, there is no reason to
restrict the sample to galaxies brighter than the completeness
limit. On the contrary, it is desirable to include faint galaxies to
improve the signal-to-noise ratio.

Note that this simplification results from the factorisation of $\mu$
in Equation~(\ref{eq:dnds_obs}) and thus only holds if the intrinsic
unlensed counts follow a power law. Note also that the likelihood
${\mathcal L}$ is independent of the completeness function
$\phi(S)$. Thus, two samples with different $\phi$'s but with the same
observed galaxy density $\tilde{n}_{0}$ would be statistically
indistinguishable, as far as the depletion analysis is concerned. This
allows us to use a power law count with a rescaled normalization
(corresponding to a constant $\phi(S)$) in the simulations, and thus
to avoid modeling the completeness function.

At the very faint end, the sample will not only be incomplete, but it
will also be contaminated by spurious sources arising from noise peaks.
These spurious sources will be uniformly distributed on the chip and
will not be affected by lensing. While this contamination could
easily be introduced in the likelihood function, it is hardly
necessary in practice. Indeed, incompleteness typically becomes
important before contamination does. As a result, a moderate magnitude
cut ensures that the contamination fraction in small. In addition, 
contamination tends to `wash out' the lensing signal. From the
point of view of detecting the lensing signal, it is therefore
conservative to ignore this effect.

\bsp

\label{lastpage}

\end{document}